\documentclass[letterpaper,11pt]{article}
\usepackage{amsmath,mathrsfs,amssymb,amsfonts,amsthm,mathtools}
\usepackage{physics,color,epsfig}
\usepackage{graphicx}
\usepackage{subfigure}
\usepackage{tabularx} 
\usepackage{booktabs}
\usepackage[table]{xcolor}
\usepackage{tikz,tikz-cd}
\usetikzlibrary{shapes.geometric, arrows, positioning}
\usepackage{empheq}

\pdfoutput=1 
\usepackage{jheppub} 
\usepackage[T1]{fontenc} 

\newcommand{\nn}{\nonumber}

\hypersetup{
    colorlinks=true,
    linkcolor=blue,
    citecolor=blue,
    urlcolor=blue
}

\title{Two-photon emission from uniform acceleration: Unruh excitation, radiative decay, and field entanglement}

\author[a]{Arash Azizi}

\affiliation[a]{{\it The Institute for Quantum Science and Engineering,
Texas A\&M University,\\ College Station, TX 77843, U.S.A.}}

\emailAdd{sazizi@tamu.edu}

\abstract{
We analyze the two-photon emission process of a uniformly accelerated Unruh-DeWitt detector interacting with a massless scalar field in Minkowski spacetime. Using second-order perturbation theory, we derive the exact final quantum state of the field and classify the emitted photon pairs according to their directional decomposition into Unruh modes: right-right (RR), left-left (LL), and mixed (RL/LR) channels. The RR and LL contributions produce photon pairs within a single Rindler wedge, while RL and LR emissions generate entanglement across wedges. The first emission arises from a counter-rotating (Unruh) excitation, followed by a rotating Wigner–Weisskopf-like decay. Our results reveal how acceleration imprints entanglement and thermal structure on the radiation, offering new insight into quantum field theory in non-inertial frames and potential applications in relativistic quantum information.}

\begin{document}

\maketitle
\flushbottom

\section{Introduction}

The Unruh effect, a cornerstone of quantum field theory (QFT) in non-inertial frames, predicts that a uniformly accelerating observer perceives the Minkowski vacuum as a thermal bath \cite{Unruh1976, Fulling1973, Davies1975}. This phenomenon is often studied using the Unruh-DeWitt (UDW) detector model—a localized two-level quantum system coupled to a quantum field \cite{Unruh1976, Einstein100, Louko2008, Colosi2009Rovelli}. UDW detectors have proven invaluable for investigating relativistic quantum phenomena, such as entanglement generation in accelerated systems \cite{Reznik2003, Salton2015harvesting, Martin-Martinez2016harvesting, Bunney2023Circularmotion, Fewster2016Louko, Zhang2020harvesting, Barman2021harvesting, Liu2022harvesting} and quantum information tasks performed in curved spacetime contexts \cite{Su2014communication, Foo2020teleportation, Tjoa2022teleportation}.

This paper extends these investigations to second-order processes by analyzing the two-photon emission from a single, uniformly accelerating UDW detector. This higher-order process provides a deeper probe into the structure of vacuum fluctuations and the dynamics of particle creation for accelerated observers. The physical mechanism involves two distinct steps: first, a counter-rotating interaction excites the detector from its ground state, a process responsible for the Unruh effect. Second, a standard rotating-wave interaction causes the detector to decay back to the ground state. This reveals an intrinsic time-ordering to the two-photon emission: an Unruh-induced excitation followed by a radiative decay, or Wigner--Weisskopf-like decay \cite{Wigner_Weisskopf1930}.

Our analysis of a second-order process for a single detector is distinct from our earlier work, which considered a system of two detectors interacting via a process that was first-order in each detector's coupling \cite{Svidzinsky21PRR, Svidzinsky21PRL}. The present work builds upon the foundational framework of Scully, Svidzinsky, and Unruh, who interpreted Unruh radiation as an entangled, two-mode squeezed state \cite{Scully2022}. While their work provided crucial physical insight, the calculation remained conceptual: it focused primarily on right-propagating modes and left the final state expressed symbolically, with the interaction strength contained within an unevaluated time integral, $\Lambda$ \cite{Scully2022}.

In contrast, our primary contribution is a complete and explicit calculation of the second-order Dyson series for this process. We derive the exact final two-photon state in analytic form for all four directional emission channels: right-right (RR), left-left (LL), right-left (RL), and left-right (LR). By fully evaluating the time integrals, we reveal the explicit resonant structure of the emission amplitudes, showing sharp peaks at the Unruh frequencies $\Omega = \pm \omega/a$, modulated by a thermal factor consistent with the Unruh temperature.

Our results demonstrate that while the RR and LL channels produce photon pairs within a single Rindler wedge, the RL and LR channels generate entangled photons across the two wedges. The generation of such non-local entanglement from a localized detector interaction highlights the underlying bipartite structure of the Minkowski vacuum itself \cite{Higuchi1992, Lin2006, Raval1996}. These findings not only deepen our understanding of observer-dependent phenomena in QFT but may also inform the development of relativistic quantum information protocols in non-inertial frames \cite{Su2014communication, Foo2020teleportation, Tjoa2022teleportation}.

\section{Field Quantization and Unruh Modes} \label{sec:field_quantization}

\begin{figure}[ht]
\centering
\includegraphics[width=.8\textwidth]{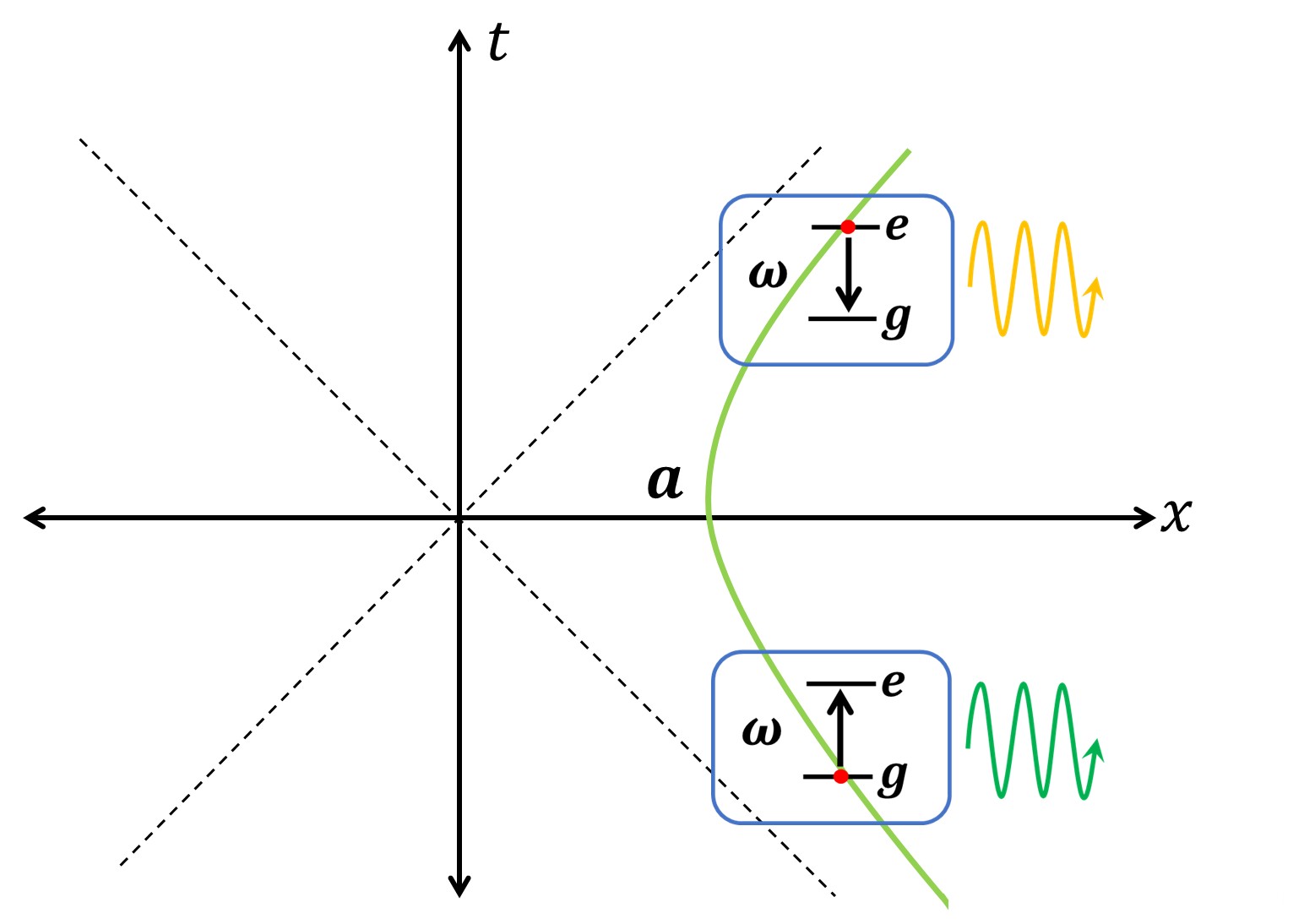}
\caption{Spacetime setup for a uniformly accelerated Unruh-DeWitt detector interacting with a scalar field in the Minkowski vacuum. The detector follows a hyperbolic trajectory with proper acceleration $a$ and energy gap $\omega$, undergoing a $g \to e \to g$ transition via Unruh excitation and radiative decay.}
\label{fig:setup}
\end{figure}

We study a uniformly accelerated Unruh-DeWitt detector, modeled as a two-level quantum system with ground state $\ket{g}$ and excited state $\ket{e}$, coupled to a massless scalar field in 1+1-dimensional Minkowski spacetime (see Figure \ref{fig:setup}). The detector follows a hyperbolic trajectory with proper acceleration $a$ and energy gap $\omega$, undergoing a $g \to e \to g$ transition that emits two photons via Unruh excitation and radiative decay. We work in light-cone coordinates $u = t - x$ and $v = t + x$, where the scalar field $\Phi$ decomposes into right-traveling ($\Phi_{\text{RTW}}(u)$) and left-traveling ($\Phi_{\text{LTW}}(v)$) components. We also set $c=1$.

The interaction Hamiltonian is
\begin{equation}
H_{\text{int}} = g \frac{\partial }{\partial \tau} \Phi(t(\tau), x(\tau))
\left( \sigma^{\dagger} e^{i \omega \tau} + \sigma e^{-i \omega \tau} \right). \label{H_int}
\end{equation}
The derivative coupling, $\frac{\partial \Phi}{\partial \tau}$, describes the detector’s interaction with the time derivative of the field along its worldline, in accordance with relativistic quantum field theory. Physically, this form of coupling is motivated by analogy with dipole-dipole interactions, where the transition rates depend on the field’s rate of change at the detector’s location. Mathematically, introducing the derivative ensures that the transformation to proper time coordinates is well-behaved: the presence of $\frac{\partial}{\partial \tau}$ precisely cancels the Jacobian arising from the coordinate change, since $du\frac{\partial}{\partial u} = d\tau \frac{\partial}{\partial \tau}$. This choice preserves the invariance and consistency of the interaction in the accelerated frame. The coupling constant $g$ quantifies interaction strength, and $\sigma, \sigma^\dagger$ are lowering and raising operators for the detector.

To analyze the interaction in the detector's accelerated frame, we expand the field in terms of Unruh modes, which are particularly suited for this geometry. The right- and left-traveling components of the scalar field, $\Phi_{\text{RTW}}(u)$ and $\Phi_{\text{LTW}}(v)$, are given by the expansions:
\begin{align}
\Phi_{\text{RTW}}(u)= \int_{-\infty}^{+\infty} d\Omega
\Bigg\{& \Big(
\theta(u) f(\Omega) u^{i\Omega}
+ \theta(-u)f(-\Omega) (-u)^{i\Omega} \Big) A_{\Omega} \nn\\
& + \Big(
\theta(u) f(\Omega) u^{-i\Omega}
+ \theta(-u) f(-\Omega) (-u)^{-i\Omega} \Big) A_{\Omega}^{\dagger}
\Bigg\}, \label{Unruh.mode.RTW}
\end{align}
and
\begin{align}
\Phi_{\text{LTW}}(v)= \int_{-\infty}^{+\infty} d\Omega
\Bigg\{& \Big(
\theta(v) f(\Omega) v^{i\Omega}
+ \theta(-v)f(-\Omega) (-v)^{i\Omega} \Big) B_{\Omega} \nn\\
& + \Big(
\theta(v) f(\Omega) v^{-i\Omega}
+ \theta(-v) f(-\Omega) (-v)^{-i\Omega} \Big) B_{\Omega}^{\dagger}
\Bigg\}, \label{Unruh.mode.LTW}
\end{align}
where $A_{\Omega}$ and $B_{\Omega}$ are the annihilation operators for right- and left-moving Unruh modes, respectively. The normalization factor
\begin{equation}
f(\Omega) = \frac{e^{- \pi \Omega / 2}}{\sqrt{8\pi \Omega \sinh(\pi \Omega)}},\label{f}
\end{equation}
is derived from the Klein-Gordon inner product \cite{UnruhWald1984}. For the photon emission process starting from the Minkowski vacuum $\ket{0_M}$, only terms with creation operators ($A^\dagger_\Omega, B^\dagger_\Omega$) will yield a non-zero contribution.

A crucial feature of this mode decomposition is that the spatial localization of an emitted photon depends on the sign of its Unruh frequency $\Omega$. The exponential factors in the normalization coefficient $f(\pm\Omega)$ cause the modes to be exponentially enhanced in one Rindler wedge while being suppressed in the causally disconnected wedge. This subtle relationship is fundamental to understanding the cross-wedge nature of the two-photon emission and is summarized in Table \ref{tab:unruh_mode_localization}.
\begin{table}[ht]
\centering
\caption{Localization of Unruh modes in the Rindler wedges. The dominant wedge for a given mode depends on the sign of its Unruh frequency, \(\Omega\), due to the exponential enhancement factor inherent in the mode functions.}
\vspace{.4cm}
\label{tab:unruh_mode_localization}
\renewcommand{\arraystretch}{1.3}
\begin{tabularx}{\textwidth}{@{} l c c X @{}}
\toprule
\textbf{\begin{tabular}[c]{@{}l@{}}Mode \\ Type\end{tabular}} & 
\textbf{\begin{tabular}[c]{@{}c@{}}Frequency \\ Sign\end{tabular}} & 
\textbf{\begin{tabular}[c]{@{}c@{}}Dominant \\ Wedge\end{tabular}} & 
\textbf{Reason} \\
\midrule
RTW (\(A_\Omega^\dagger\)) & \(\Omega > 0\) & Right & The \(\theta(-u)\) term is active, with its coefficient \(f(-\Omega) \propto e^{\pi \Omega / 2}\) providing exponential enhancement. \\
\rowcolor{gray!15}
RTW (\(A_\Omega^\dagger\)) & \(\Omega < 0\) & Left & The \(\theta(u)\) term is active, with its coefficient \(f(\Omega) \propto e^{\pi |\Omega| / 2}\) providing enhancement. \\
\addlinespace
LTW (\(B_\Omega^\dagger\)) & \(\Omega > 0\) & Left & The \(\theta(-v)\) term is active, with its coefficient \(f(-\Omega) \propto e^{\pi \Omega / 2}\) providing enhancement. \\
\rowcolor{gray!15}
LTW (\(B_\Omega^\dagger\)) & \(\Omega < 0\) & Right & The \(\theta(v)\) term is active, with its coefficient \(f(\Omega) \propto e^{\pi |\Omega| / 2}\) providing enhancement. \\
\bottomrule
\end{tabularx}
\end{table}
For the photon emission process starting from the Minkowski vacuum $\ket{0_M}$, only terms with creation operators ($A^\dagger_\Omega, B^\dagger_\Omega$) will yield a non-zero contribution.

The detector moves along the trajectory $u = -\frac{1}{a} e^{-a\tau}$ and $v = \frac{1}{a} e^{a\tau}$. Expressing the field operators in the detector's proper time $\tau$ within the right Rindler wedge ($u < 0, v > 0$), we get:
\begin{align}
\Phi_{\text{RTW}}(\tau)
&= \int_{-\infty}^{+\infty} d\Omega f(-\Omega)\Big(
a^{-i\Omega} e^{-i a \Omega \tau} A_\Omega
+ a^{i\Omega} e^{i a \Omega \tau} A_\Omega^\dagger
\Big), \nn \\
\Phi_{\text{LTW}}(\tau)
&= \int_{-\infty}^{+\infty} d\Omega
f(\Omega)\Big(
a^{-i\Omega} e^{i a \Omega \tau} B_\Omega
+ a^{i\Omega} e^{-i a \Omega \tau} B_\Omega^\dagger \Big). \label{Unruh.mode.tau}
\end{align}
The Jacobian relations
\begin{equation}
du \frac{\partial}{\partial u} = d\tau \frac{\partial}{\partial \tau}, \quad dv \frac{\partial}{\partial v} = d\tau \frac{\partial}{\partial \tau},
\end{equation}
transform the Dyson series into proper-time integrals.
\vspace{.3cm}

To appreciate the physical consequences of this mode structure before proceeding to the second-order calculation, it is instructive to briefly review the first-order Unruh-Wald effect. In this process, the detector is excited from $\ket{g}$ to $\ket{e}$, and the final state is given by the first-order Dyson series:
\begin{align}
\ket{\Psi_f} = \frac{-i}{\hbar}
\int_{-\infty}^{+\infty} d\tau g \frac{\partial \Phi}{\partial \tau} \sigma^\dagger e^{i \omega \tau}
\ket{0_M} \ket{g}.
\label{psi_f-tau_first.1}
\end{align}
Substituting the right-traveling field from Eq.~\eqref{Unruh.mode.tau} yields:
\begin{align}
\ket{\Psi_f} =& \frac{-i}{\hbar}
\int_{-\infty}^{+\infty} d\tau g \int_{-\infty}^{+\infty} d\Omega f(-\Omega) a^{i\Omega}
\frac{\partial }{\partial \tau} e^{i a \Omega \tau} 
\,A_\Omega^\dagger e^{i \omega \tau}
\ket{0_M} \ket{e} \nn\\
=& \frac{2\pi g}{\hbar}
\int_{-\infty}^{+\infty}
d\Omega \Omega f(-\Omega) a^{1+i\Omega} A_\Omega^\dagger
\delta \big( a\Omega+\omega \big)
\ket{0_M} \ket{e} \nn\\
=& \frac{2\pi g}{\hbar}
(-\frac{\omega}{a}) f\big(\frac{\omega}{a}\big) a^{-i\frac{\omega}{a}}
A_{-\frac{\omega}{a}}^\dagger \ket{0_M} \ket{e}.
\label{psi_f-tau_first.2}
\end{align}
The delta function enforces the resonance condition $\Omega=-\frac{\omega}{a}$. This result reveals a key insight from Unruh and Wald: the detector, located in the right Rindler wedge, becomes excited by emitting a photon into a negative-frequency Unruh mode. As dictated by the structure of the mode functions in Eq.~\eqref{Unruh.mode.RTW}, this negative-frequency excitation ($A_{-\omega/a}^\dagger$) predominantly populates the left Rindler wedge ($\theta(u)$ region), which is causally disconnected from the detector. This non-local emission is a direct consequence of the entangled structure of the Minkowski vacuum and is foundational to the two-photon process we analyze next.

\section{Time Evolution} \label{sec. time evol}

We calculate the final two-photon state by beginning with the detector in its ground state $\ket{g}$ and the field in the Minkowski vacuum $\ket{0_M}$. The state evolution is governed by the second-order Dyson series, where the first-order term corresponds to the detector's excitation and the second-order term captures the full $g \to e \to g$ emission process. This second-order contribution to the final state is:
\begin{align}
\ket{\Psi_f} = \left( \frac{-i}{\hbar} \right)^2
\int_{-\infty}^{+\infty} d\tau g \frac{\partial \Phi}{\partial \tau} \sigma e^{-i \omega \tau}
\int_{-\infty}^{\tau} d\tau' g \frac{\partial \Phi}{\partial \tau'} \sigma^{\dagger} e^{i \omega \tau'}
\ket{0_M} \ket{g}.
\label{psi_f-tau}
\end{align}
In this expression, the detector is first excited at proper time $\tau'$ (governed by $\sigma^{\dagger} e^{i \omega \tau'}$) and later de-excites at time $\tau$ (governed by $\sigma e^{-i \omega \tau}$).

To evaluate this integral, we substitute the full field operator $\Phi = \Phi_{\text{RTW}} + \Phi_{\text{LTW}}$ using the Unruh mode expansion from Eq.~\eqref{Unruh.mode.tau}. Since the field starts in the vacuum, only the creation operator terms ($A_{\Omega}^\dagger, B_{\Omega}^\dagger$) will give a non-zero result:
\begin{align}
&\ket{\Psi_f} = \label{psi_f.2}\\
&
\left( \frac{-ig}{\hbar} \right)^2
\int_{-\infty}^{+\infty} d\tau
\int_{-\infty}^{+\infty} d\Omega \Bigg\{
f(-\Omega) \frac{\partial }{\partial \tau}  e^{i a \Omega \tau}
a^{i\Omega} A^\dagger_{\Omega}
 e^{-i \omega \tau}
 + f(\Omega) \frac{\partial }{\partial \tau}  e^{-i a \Omega \tau}
a^{i\Omega} B^\dagger_{\Omega}
 e^{-i \omega \tau} \Bigg\}  \nn\\
&\times
\int_{-\infty}^{\tau} d\tau'
\int_{-\infty}^{+\infty} d\Omega' \Bigg\{
f(-\Omega') \frac{\partial }{\partial \tau'}  e^{i a \Omega' \tau'}
a^{i\Omega'} A^\dagger_{\Omega'}
 e^{i \omega \tau'}
 + f(\Omega') \frac{\partial }{\partial \tau'}  e^{-i a \Omega' \tau'}
a^{i\Omega'} B^\dagger_{\Omega'}
 e^{i \omega \tau'} \Bigg\}
\ket{0_M} \ket{g}.  \nn
\end{align}
Next, we evaluate the derivatives with respect to proper time. The term $\frac{\partial }{\partial \tau} e^{\pm i a \Omega \tau}$ introduces a factor of $\pm i a \Omega$. Applying this to both the $\tau$ and $\tau'$ integrals and combining constant factors leads to:
\begin{align}
&\ket{\Psi_f} = \label{psi_f.3}\\
&
\left( \frac{g}{\hbar} \right)^2
\int_{-\infty}^{+\infty} d\tau
\int_{-\infty}^{+\infty} d\Omega \Omega \Bigg\{
f(-\Omega) e^{i a \Omega \tau}
a^{1+i\Omega} A^\dagger_{\Omega}
 e^{-i \omega \tau}
 - f(\Omega)  e^{-i a \Omega \tau}
a^{1+i\Omega} B^\dagger_{\Omega}
 e^{-i \omega \tau} \Bigg\}  \nn\\
&\times
\int_{-\infty}^{\tau} d\tau'
\int_{-\infty}^{+\infty} d\Omega'\Omega' \Bigg\{
f(-\Omega') e^{i a \Omega' \tau'}
a^{1+i\Omega'} A^\dagger_{\Omega'}
 e^{i \omega \tau'}
 - f(\Omega')  e^{-i a \Omega' \tau'}
a^{1+i\Omega'} B^\dagger_{\Omega'}
 e^{i \omega \tau'} \Bigg\}
\ket{0_M} \ket{g}.  \nn
\end{align}
The inner integral over $\tau'$ is a standard Fourier integral, which evaluates to:
\begin{align}
&\ket{\Psi_f} = \label{psi_f.4}\\
&
\left( \frac{g}{\hbar} \right)^2
\int_{-\infty}^{+\infty} d\tau
\int_{-\infty}^{+\infty} d\Omega \Omega \Bigg\{
f(-\Omega) e^{i a \Omega \tau}
a^{1+i\Omega} A^\dagger_{\Omega}
 e^{-i \omega \tau}
 - f(\Omega)  e^{-i a \Omega \tau}
a^{1+i\Omega} B^\dagger_{\Omega}
 e^{-i \omega \tau} \Bigg\}  \nn\\
&\times
\int_{-\infty}^{+\infty} d\Omega'\Omega' \Bigg\{
f(-\Omega') \frac{e^{i \big(a \Omega' +\omega\big)  \tau}}
{i\big (a \Omega' +\omega\big)}
a^{1+i\Omega'} A^\dagger_{\Omega'}
 - f(\Omega')  \frac{e^{i \big(-a \Omega' +\omega\big)  \tau}}
{i \big(-a \Omega' +\omega\big)}
a^{1+i\Omega'} B^\dagger_{\Omega'}
 \Bigg\}
\ket{0_M} \ket{g}.  \nn
\end{align}
Performing the final integral over $\tau$ yields a Dirac delta function, which enforces energy conservation for the overall two-photon process and correlates the Unruh frequencies of the emitted photons.
\begin{align}
\ket{\Psi_f} =
\frac{2\pi g^2}{\hbar^2}
\int_{-\infty}^{+\infty} d\Omega \Omega
\int_{-\infty}^{+\infty}& d\Omega'\Omega' \label{psi_f.5}\\
\times \Bigg\{&
f(-\Omega) f(-\Omega') \frac{a^{2+i\big(\Omega+\Omega'\big)}}
{i\big(a \Omega' +\omega\big)}
\delta \Big( a \big(\Omega+\Omega'\big) \Big)
 A^\dagger_{\Omega}  A^\dagger_{\Omega'} \nn\\
&
- f(-\Omega) f(\Omega') \frac{a^{2+i\big(\Omega+\Omega'\big)}}
{i\big(-a \Omega' +\omega\big)}
\delta \Big( a \big(\Omega-\Omega'\big) \Big)
 A^\dagger_{\Omega}  B^\dagger_{\Omega'} \nn\\
&
- f(\Omega) f(-\Omega') \frac{a^{2+i\big(\Omega+\Omega'\big)}}
{i\big(a \Omega' +\omega\big)}
\delta \Big( a \big(\Omega-\Omega'\big) \Big)
 B^\dagger_{\Omega} A^\dagger_{\Omega'}   \nn\\
&
+ f(\Omega) f(\Omega') \frac{a^{2+i\big(\Omega+\Omega'\big)}}
{i\big(-a \Omega' +\omega\big)}
\delta \Big( a \big(\Omega+\Omega'\big) \Big)
 B^\dagger_{\Omega}  B^\dagger_{\Omega'} \Bigg\}
\ket{0_M} \ket{g}.  \nn
\end{align}
The four terms in the brackets correspond to the four possible emission channels: RR ($A^\dagger A^\dagger$), RL ($A^\dagger B^\dagger$), LR ($B^\dagger A^\dagger$), and LL ($B^\dagger B^\dagger$). To obtain the final state, we evaluate the delta functions, which reduces the double integral to a single integral over $\Omega$. Using the property $\delta(ax) = \frac{1}{|a|} \delta(x)$ and the normalization product from Eq.~\eqref{f}, we arrive at the final analytical form of the two-photon state:
\begin{align}
 \boxed{   \ket{\Psi_f} =
\frac{ i g^2}{4\hbar^2}
\int_{-\infty}^{+\infty} d\Omega
\frac{1}{\sinh(\pi \Omega)} 
\Bigg\{ \frac{\Omega}{\frac{\omega}{a} - \Omega}
 A^\dagger_{\Omega}  A^\dagger_{-\Omega}
 + \frac{\Omega}{\frac{\omega}{a} + \Omega}
 B^\dagger_{\Omega}  B^\dagger_{-\Omega}
 + \frac{2\frac{\omega}{a}\,\Omega\,  a^{2i\Omega} }{(\frac{\omega}{a})^2 - \Omega^2}
 A^\dagger_{\Omega}  B^\dagger_{\Omega} \Bigg\}
\ket{0_M} \ket{g}. } \label{psi_f.6}
\end{align}

\section{Discussion and Conclusion} \label{sec:discussion_conclusion}

This work presents a second-order Dyson series analysis of two-photon emission from a uniformly accelerated Unruh-DeWitt detector coupled to a massless scalar field in (1+1)-dimensional Minkowski spacetime. As derived in Section \ref{sec. time evol}, the complete two-photon final state (Eq.~\eqref{psi_f.6}) decomposes into three directional channels: right-right (RR), left-left (LL), and mixed (RL+LR), each revealing the Minkowski vacuum's bipartite entanglement across Rindler wedges.

The labels RTW (right-traveling wave, dependent on $u = t - x$) and LTW (left-traveling wave, dependent on $v = t + x$) denote propagation direction, not spatial location in the Rindler wedges. The photon's predominant support in the right ($u < 0$, $v > 0$) or left ($u > 0$, $v < 0$) wedge depends on the Unruh frequency $\Omega$: positive $\Omega > 0$ modes are exponentially localized in one wedge (e.g., right for RTW, left for LTW), while negative $\Omega < 0$ modes are localized in the opposite wedge, due to the mode functions' structure (Eqs.~\eqref{Unruh.mode.RTW}--\eqref{Unruh.mode.LTW}) and normalization factors $f(\pm \Omega) \sim e^{\mp \pi \Omega / 2}$ introducing exponential enhancement/suppression \cite{Unruh1976, Higuchi1992}.

The Minkowski vacuum $\ket{0_M}$ is a two-mode squeezed state entangled between the left (L) and right (R) Rindler wedges:
\begin{equation}
\ket{0_M} = \prod_k \sqrt{1 - e^{-2\pi k / a}} \sum_{n=0}^\infty e^{-\pi n k / a} \ket{n_k}_R \ket{n_k}_L,
\end{equation}
correlating modes across causally disconnected wedges. In our process, the detector, confined to the right wedge, probes this entanglement via a two-step mechanism. First, a counter-rotating interaction excites the detector. This excitation serves as a local probe of the vacuum's non-local entanglement. While the resulting detector-field state after this initial step is a product state (Sec.~\ref{sec:field_quantization}, Eq.~\eqref{psi_f-tau_first.2}), the entanglement is witnessed by the non-local character of the particle creation: the detector's excitation in the right wedge is inextricably linked to the emission of a negative-frequency Unruh photon into the causally disconnected left wedge. This excitation into the left wedge is only possible due to the vacuum’s correlations; a separable vacuum (e.g., $\ket{0_R} \otimes \ket{0_L}$) would prohibit thermal excitation \cite{UnruhWald1984}. Second, a rotating decay releases this energy as real photons, effectively converting the probed vacuum correlations into the final entangled two-photon state.

All channels produce cross-wedge entangled pairs: RR ($A_\Omega^\dagger A_{-\Omega}^\dagger$, resonance $\Omega = \omega/a > 0$) pairs a right-moving photon in the right wedge ($\Omega > 0$) with one in the left wedge ($-\Omega < 0$); LL ($B_\Omega^\dagger B_{-\Omega}^\dagger$, $\Omega = -\omega/a < 0$) pairs a left-moving photon in the right wedge with one in the left; RL+LR ($A_\Omega^\dagger B_\Omega^\dagger$, $\Omega = \pm \omega/a$) pairs right- and left-moving photons across wedges. The thermal factor $1/\sinh(\pi \Omega)$ (reflecting Unruh temperature $T_U = a/2\pi$) and phase $a^{2i\Omega}$ (from trajectory scaling) in RL+LR are imprints of this vacuum structure. If entanglement were solely detector-generated (e.g., inertial parametric down-conversion), we would expect no thermal spectrum or cross-wedge emissions, and the process would persist as $a \to 0$, contrary to our results where peaks shift to $\Omega \to \infty$ and vanish \cite{Scully2022}. Thus, all channels witness the vacuum's bipartite entanglement, akin to entanglement harvesting \cite{Reznik2003, Salton2015harvesting}.

\begin{figure}[ht]
\centering
\includegraphics[width=\textwidth]{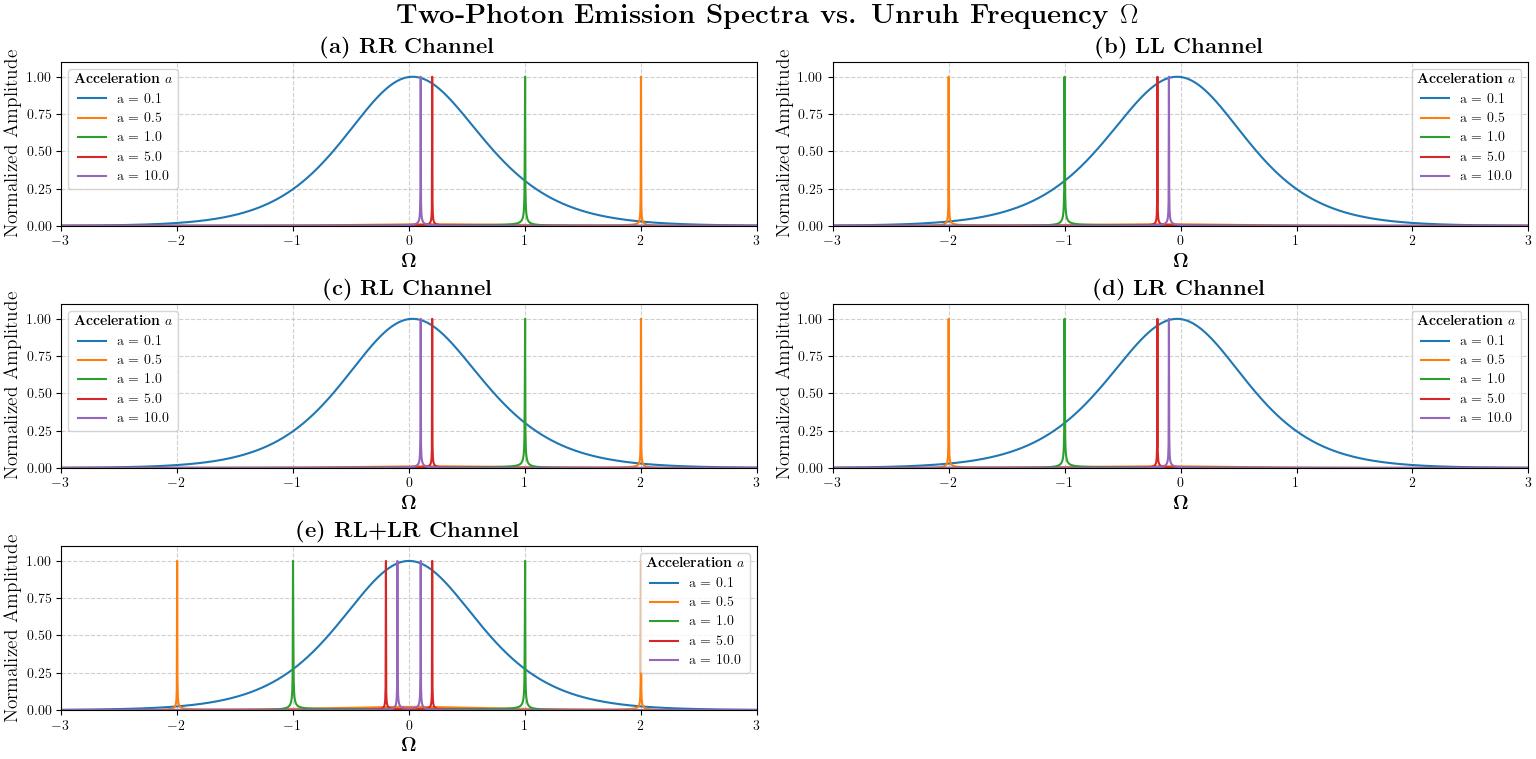}
\caption{Normalized amplitude of two-photon emission vs. Unruh frequency $\Omega \in [-3, 3]$, for detector energy gap $\omega = 1$ and accelerations $a = 0.1, 0.5, 1.0, 5.0, 10.0$. (a) RR channel, with approximate resonances at $\Omega = \omega/a = 1/a = [10.0, 2.0, 1.0, 0.2, 0.1]$; the $a = 0.1$ peak ($\Omega = 10$) is outside the range, driven by rotating second emission following counter-rotating excitation via vacuum fluctuations. (b) LL channel, with approximate resonances at $\Omega = -\omega/a = [-10.0, -2.0, -1.0, -0.2, -0.1]$; $a = 0.1$ peak outside. (c) RL channel, with resonances at $\Omega = 1/a$. (d) LR channel, with resonances at $\Omega = -1/a$. (e) RL+LR channel, with symmetric resonances at $\Omega = \pm \omega/a = \pm 1/a$, reflecting entanglement across Rindler wedges. The $\sinh(\pi\Omega)$ factor suppresses high frequencies, broadening peaks, while raw amplitudes vary with $a$, and normalization highlights peak shifts within $[-3, 3]$.}
\label{fig:emission_spectra}
\end{figure}

The directional and entangled structure is summarized in Table \ref{tab:channel_summary} and visualized in Figure \ref{fig:emission_spectra}. The counter-rotating process converts vacuum correlations into entangled photons, bridging quantum and classical regimes \cite{Wald1994}. These results may suggest potential observables, such as photon coincidence measurements, for theoretical studies of Unruh radiation in high-acceleration systems and contribute to understanding relativistic quantum information protocols in non-inertial frames, such as entanglement harvesting \cite{Reznik2003}. Future work could extend this to curved spacetimes or higher-order perturbations to explore multipartite entanglement, leveraging the $\chi$-based formalism (Appendix \ref{app. chi}).

\begin{table}[ht]
\centering
\setlength{\tabcolsep}{10pt} 
\renewcommand{\arraystretch}{1.4} 
\caption{Summary of two-photon emission channels. Note that all channels result in cross-wedge entanglement, connecting the two causally disconnected Rindler wedges.}
\vspace{.4cm}
\label{tab:channel_summary}
\begin{tabular}{|c|c|c|p{6cm}|}
\hline
\textbf{Channel} & \textbf{Operators} & \textbf{Resonance} & \textbf{Key Features} \\
\hline
RR & \(A_{\Omega}^\dagger A_{-\Omega}^\dagger\) & \(\Omega = \omega/a\) & A pair of right-moving photons entangled across opposite wedges. \\
\rowcolor{gray!20}
LL & \(B_{\Omega}^\dagger B_{-\Omega}^\dagger\) & \(\Omega = -\omega/a\) & A pair of left-moving photons entangled across opposite wedges. \\
RL+LR & \(A_{\Omega}^\dagger B_{\Omega}^\dagger\) & \(\Omega = \pm \omega/a\) & Entanglement between right- and left-moving photons with symmetric resonances. \\
\hline
\end{tabular}
\end{table}

\section*{Acknowledgments}
I am grateful to Marlan Scully, Anatoly Svidzinsky, and Bill Unruh for illuminating discussions. I would also like to thank the anonymous referee for their insightful comments, which significantly improved the clarity and focus of this manuscript. This work was supported by the Robert A. Welch Foundation (Grant No. A-1261) and the National Science Foundation (Grant No. PHY-2013771).
\appendix 

\section{Unified formalism} \label{app. chi}

Here, we present a unified derivation for all four emission channels by employing a direction parameter $\chi = \pm 1$. This parameter allows us to treat right-traveling waves (RTW, $\chi=1$, coordinate $u=t-x$) and left-traveling waves (LTW, $\chi=-1$, coordinate $v=t+x$) within a single formalism. We can unify the light-cone coordinates by defining a single coordinate $w = t - \chi x$. This unified coordinate $w$ represents the right-traveling coordinate $u=t-x$ when the direction parameter $\chi=1$, and the left-traveling coordinate $v=t+x$ when $\chi=-1$. The field $\Phi_\chi(w)$ is expanded in terms of Unruh modes as:
\begin{align}
\Phi_{\chi}(w)= \int_{-\infty}^{+\infty} d\Omega
\Bigg\{& \Big(
\theta(w) f(\Omega) w^{i\Omega}
+ \theta(-w) f(-\Omega) (-w)^{i\Omega} \Big) A_{\chi,\Omega} \nn \\
&+ \Big( \theta(w) f(\Omega) w^{-i\Omega}
+ \theta(-w) f(-\Omega) (-w)^{-i\Omega} \Big) A_{\chi,\Omega}^{\dagger}
\Bigg\},  \label{Unruh.mode}
\end{align}
where the normalization factor $f(\Omega)$ is given in Eq. \eqref{f}. The operators $A_{\chi,\Omega}$ and $A_{\chi,\Omega}^\dagger$ satisfy $[A_{\chi,\Omega}, A_{\chi',\Omega'}^\dagger] = \delta_{\chi\chi'} \delta(\Omega - \Omega')$, with $A_{1,\Omega} \equiv A_\Omega$ (RTW) and $A_{-1,\Omega} \equiv B_\Omega$ (LTW).

\vspace{.3cm}
For a detector in the right Rindler wedge, its trajectory is given by $-a u = e^{-a\tau}$ and $a v = e^{a\tau}$. This can be written compactly as $-\chi a w = e^{-\chi a \tau}$. Substituting this into the mode expansion and evaluating in the right wedge, the field in terms of the detector's proper time $\tau$ becomes:
\begin{align}
\Phi_{\chi}(\tau)
&= \int_{-\infty}^{+\infty} d\Omega f(-\chi\Omega)\Big(
a^{-i\Omega} e^{-i \chi a \Omega \tau} A_{\chi,\Omega}
+ a^{i\Omega} e^{i \chi a \Omega \tau} A_{\chi,\Omega}^\dagger
\Big). \label{Unruh.mode.chi}
\end{align}
The state for a given channel, specified by the pair $(\chi, \chi')$, is calculated using the second-order Dyson series:
\begin{align}
\ket{\Psi_f}_{\chi \chi'} = \left( \frac{-i}{\hbar} \right)^2
\int_{-\infty}^{+\infty} d\tau g
\frac{\partial }{\partial \tau} \Phi_{\chi}
\sigma e^{-i \omega \tau}
\int_{-\infty}^{\tau} d\tau' g
\frac{\partial }{\partial \tau'} \Phi_{\chi'}
\sigma^{\dagger} e^{i \omega \tau'}
\ket{0_M} \ket{g}.
\label{psi_f-chichi'.1}
\end{align}
Substituting the field expansion from Eq. \eqref{Unruh.mode.chi} and retaining only the creation operator terms gives:
\begin{align}
\ket{\Psi_f}_{\chi \chi'} =& \left( \frac{-i}{\hbar} \right)^2
\int_{-\infty}^{+\infty} d\tau g
\int_{-\infty}^{+\infty} d\Omega f(-\chi\Omega) \frac{\partial }{\partial \tau}  e^{i \chi a \Omega \tau}
a^{i\Omega} A^\dagger_{\chi,\Omega}
 e^{-i \omega \tau}  \nn\\
&\times
\int_{-\infty}^{\tau} d\tau' g
\int_{-\infty}^{+\infty} d\Omega' f(-\chi'\Omega')
\frac{\partial }{\partial \tau'}
e^{i \chi' a \Omega' \tau'}
a^{i\Omega'} A^\dagger_{\chi',\Omega'}
 e^{i \omega \tau'}
\ket{0_M} \ket{g}.
\label{psi_f-chichi'.2}
\end{align}
Integrating over \(\tau'\) gives
\begin{align}
\ket{\Psi_f}_{\chi\chi'} =& \chi\chi' \frac{g^2}{\hbar^2} 
\int_{-\infty}^{+\infty} d\tau 
\int_{-\infty}^{+\infty} d\Omega \, \Omega
f(\chi\Omega)\, e^{i \chi a \Omega \tau} 
a^{1+i\Omega}  \,
 e^{-i \omega \tau}  \nn\\
&\times
\int_{-\infty}^{+\infty} d\Omega' \, \Omega' f(\chi'\Omega')\, 
\frac{e^{i \big(\chi' a \Omega'+\omega\big)\tau}}
{i \big( \chi' a \Omega'+\omega \big)} 
a^{1+i\Omega'} A^\dagger_{\chi,\Omega} A^\dagger_{\chi',\Omega'} \,
 \ket{0_M} \ket{g}. 
\label{psi_f-chichi'.4}
\end{align}
Integrating over \(\tau\) produces a Dirac delta function:
\begin{align}
\ket{\Psi_f}_{\chi\chi'} = \chi\chi' \frac{2\pi g^2}{\hbar^2} 
\int_{-\infty}^{+\infty}& d\Omega \, \Omega
f(\chi\Omega)\,
\int_{-\infty}^{+\infty} d\Omega' \, \Omega' f(\chi'\Omega')\,
a^{1+i\Omega} 
   \nn\\
&\times 
\frac{1}{i \big(\chi' a  \Omega'+\omega\big)} 
a^{1+i\Omega'}   \delta \big(a (\chi\Omega + \chi' \Omega') \big)
A^\dagger_{\chi,\Omega} A^\dagger_{\chi',\Omega'} \,
 \ket{0_M} \ket{g}. 
\label{psi_f-chichi'.5}
\end{align}
The delta function simplifies as
\begin{align}
 \delta \big(a (\chi\Omega + \chi' \Omega') \big)=\frac{1}{a}
 \delta \big(\chi  \Omega + \chi'  \Omega') \big)=
 \delta \big( \chi \chi' \Omega +  \Omega' \big).
\end{align}
This reduces the double integral to
\begin{align}
\ket{\Psi_f}_{\chi\chi'} =& \chi\chi' \frac{2\pi g^2}{\hbar^2} 
\int_{-\infty}^{+\infty} d\Omega \,(-\chi\chi')\, \Omega^2
f(\chi\Omega)f(-\chi\Omega)\,
 \frac{a^{1+i\Omega(1-\chi\chi')}}{i \big(-\chi a \Omega+\omega\big)} 
  A^\dagger_{\chi,\Omega} A^\dagger_{\chi', -\chi\chi'\Omega} \,
 \ket{0_M} \ket{g}. 
\label{psi_f-chichi'.6}
\end{align}
Using Eq. \eqref{f}, the normalization product becomes
\begin{align}
f(\chi\Omega)f(-\chi\Omega) = \frac{1}{8\pi \Omega \sinh(\pi\Omega)}. \label{f_product}
\end{align}
Consequently arrive at the final unified expression for the two-photon state:
\begin{align}
\boxed{\quad \ket{\Psi_f}_{\chi\chi'}
= \frac{i g^2}{4\hbar^2} \int_{-\infty}^{+\infty} d\Omega
\frac{\Omega}{\left(\frac{\omega}{a} - \chi \Omega\right)}
\frac{1}{\sinh(\pi\Omega)} a^{i\Omega(1-\chi\chi')}
A^\dagger_{\chi,\Omega}
A^\dagger_{\chi', -\chi\chi'\Omega}
\ket{0_M}\ket{g}\,. \quad} \label{psi_f_final}
\end{align}
This compact state describes two photons with correlated Unruh frequencies. The specific choice of $\chi, \chi' = \pm 1$ generates the four distinct channels (RR, LL, RL, LR) discussed in the main text, with the RR and LL channels producing intra-wedge pairs and the RL and LR channels generating entangled pairs across Rindler wedges, reflecting the Minkowski vacuum’s nonlocal structure.

\bibliographystyle{jhep}
\bibliography{UnruhRef}
\end{document}